# Quark Confinement and the Fractional Quantum Hall Effect

Hai-Jun Wang

*Center for Theoretical Physics, Jilin University, Changchun 130023, China*

W. T. Geng

*Department of Materials Science & Engineering and Department of Physics, University of Science and Technology Beijing, Beijing 100083, China*

**Abstract**

Working in the physics of Wilson factor and Aharonov-Bohm effect, we find in the fluxtube-quark system the topology of a baryon consisting three heavy flavor quarks resembles that of the fractional quantum Hall effect (FQHE) in condensed matter. This similarity yields the result that the constituent quarks of baryon have the "filling factor" $1/3$, thus the previous conjecture that quark confinement is a correlation effect was confirmed. Moreover, by deriving a Hamiltonian of the system analogous to that of FQHE, we predict an energy gap for the ground state of a heavy three-quark system.

PACS numbers: 12.38.Aw, 12.39.Jh

Understanding confinement of quarks in hadrons remains a challenge, notwithstanding a mature quantum chromodynamics (QCD). The attempts to explain why quarks are confined in hadrons started from t'Hooft's work a quarter of century ago,[1] and various mechanisms within QCD have been proposed ever since. [2,3,4] In no mechanisms can we escape the task to reduce the effective degrees of freedom of QCD using some gauge fixing conditions to separate the physical and unphysical parts. Unlike QED, in a non-Abelian gauge field like QCD, the physical results can be dependent on the gauge conditions and in general the ghost fields cannot be decoupled with the vector fields. To understand quark confinement, QCD has been formulized under several gauges. On the maximum Abelian gauge fixing, it is proved that the appearance of magnetic monopoles[1] can provid a picture to explain why $q\bar{q}$ is confined[2]: The binding of a quark and an anti-quark can be understood as the result of condensation of monopoles that squeeze the force between a quark and an anti-quark to a string, known as linear confining potential. The other mostly considered gauge is the Coulomb gauge, [3,4] in which the contribution from ghosts lies in the resultant potential. There are still other gauges used to reduce the gauge symmetry of QCD in discussing the confinement of quarks, such as Landau gauge,[5,6] temporal gauge,[7] and axial gauge condition.[8] More directly, some authors lower the dimension of QCD to $QCD_2$[9] or $QCD_3$[10] to reduce the number of gauge degrees and then come back to compare results with $QCD_4$. But in all of these approaches, we encounter difficulties to derive the interacting potential in the presence of ghost-vector coupling.

There have been quite a few works[11][12][13] from the angle of condensed matter physics conjecturing the properties of confinement of strong interaction, but scientists still lack solid proofs from the viewpoint of particle physics or chromo-dynamics supporting the conjectures. If the efforts of correlating these two sides can be realized, the table-top experiments of condensed matters can be used to test speculations on hadron physics——where we have encounter many difficulties originated from the confining quarks and hence the nonperturbative properties.

Here we propose a potential model that does not fit into the paradigms of the quark--anti-quark confinement, based on which we take topological properties to explain the general quark-confinement in hadrons. Inspired by Laughlin's statement[14] that quark confinement might come from collective excitation, we have succeeded in choosing a reasonable gauge condition resembling the physical conditions for the FQHE to appear (named FQHE gauge hereafter). We find that under the FQHE gauge, the topology of the three-quark color-singlet system is equivalent to the topology of the FQHE with a filling factor of 1/3. In other words, imposing the FQHE to quark system would induce naturally the topology of the FQHE. Together with the heavy-quark approximation, the FQHE gauge makes the complex interaction between ghosts and vector bosons irrelevant to quark confinement.

The topology of FQHE[15] depends on that of Aharonov-Bohm effect[16], and thus on closed Wilson lines (named as nonintegrable phase factor in some other references). Since QCD can be equivalently formulated as a theory of Wilson lines, many authors started with Wilson loops to derive the strong interaction between quarks [17] [18] [19] [20]. An excellent treatment in the derivation is to study systems consisting heavy flavor quarks ($m \gg \Lambda_{QCD}$). These systems are near perturbative region and the non-relativistic Schrödinger equation is a fairly good approximation at leading order. In this paper, without affecting conclusions, we qualitatively make the quarks in a baryon have the same masses.

Before studying the topology of QCD, we first reduce the degrees of freedom of three-quark baryon system. Our model is a hybrid scenario of combining the covariant gauge condition $\partial_\mu A^\mu = 0$ and the situation that FQHE happens (FQHE gauge). The FQHE takes place in a 2-dimension system with a vertical magnetic field penetrating through. Since our focus here is the origination and degeneration of the states in energy bands, the electric field perpendicular to the magnetic field in FQHE is neglected in our discussion. The introduction of such a gauge scheme prompts us to lower the dimension of interactions in QCD down to 2. As shown in the following, however, the field $A_\mu$ here is generated in a way different from $QCD_2$. In a three-quark color-singlet system, one quark (hereafter referred to as the *non-local quark*) wanders in the field induced by the other two quarks (hereafter referred to as the *di-quark*). Recently, Alexandrou *et al*. have evidenced the existence of stable *di-quarks* in the lattice calculation.[21] The magnetic part of the field induced by *di-quark* penetrates (not vertically, in contrast to the case in FQHE) instantaneously through a plane, and the *non-local quark* locates in this very plane to meet the conservation of momentum and angular momentum. The dynamics of the *non-local quark* is determined by the normal Lagrangian in QCD:

$$\mathcal{L} = \bar{\psi}(i\mathcal{D})\psi - \frac{1}{4}(F^i_{\mu\nu})^2 - m\bar{\psi}\psi \ . \qquad (1)$$

Here $D_\mu = \partial_\mu - i g A^a_\mu \frac{\lambda^a}{2}$ [$\lambda^a$ is Gellmann matrices] and $F^i_{\mu\nu}$ is defined as

$$F_{\mu\nu} = F^a_{\mu\nu}\frac{\lambda^a}{2} = \partial_\mu A^a_\nu \frac{\lambda^a}{2} - \partial_\nu A^a_\mu \frac{\lambda^a}{2} - i g [A^a_\mu \frac{\lambda^a}{2}, A^b_\nu \frac{\lambda^b}{2}], \qquad (2)$$

where $A_\mu = A^a_\mu \frac{\lambda^a}{2}$ is the vector field induced by the *di-quark*. Gluons are assumed to be massless here. There are two reasons for us to choose such an ordinary Lagrangian for the *non-local* quark. First, the *di-quark* interacts as a whole with the *non-local* quark. Second, under the condition of color-singlet for a three-quark system, the *di-quark* behaves, in the sense of color, like an anti-quark of the *non-local quark*.

The covariant gauge fixing $\partial_\mu A^\mu = 0$ is imposed before the *non-local quark* is required to be in a plane. To simulate the situation of FQHE, we further idealize the scalar component $A_0$ in $A_\mu = (A_0, \vec{A})$ to be nontrivial only at infinite point or the interaction core (Pauli principle) of the plane. Since we discuss only heavy quarks ($m \gg \Lambda_{QCD}$) that need not wander too far away from the center of the field induced by the *di-quark* to meet the infrared requirement, $A_0$ is taken as zero in what follows. With $\partial_\mu A^\mu = 0$ and $A_0 \approx 0$, the spatial degrees of freedom of $A_\mu$ is reduced to two.[22] Under the heavy-quark approximation, tree level diagram plus renormalization can well describe the scattering amplitude and the changes of wave functions. The ghosts' contribution to the wave function, which appears only in higher order corrections and hence in renormalization, may not be relevant in our case because the renormalization will not affect topology.

Now, let's turn to the topology of QCD. With reduced degrees of freedom, we can write the field strength Eq. (2) of the *di-quark* as

$$\vec{B} = \vec{\nabla} \times \vec{A} - ig\, \vec{A} \times \vec{A} \;, \qquad (3)$$

We note that $\vec{A}$ lies in the plane, and the second term prevent $\vec{B}$ to be perpendicular to the plane.[23] Since the quark move in the plane within a restricted region, its paths (wave) form loops. It therefore justifies our use of the form of closed Wilson line in non-Abelian fields. To analogize the FQHE, let's study the Wilson line along an infinitesimal closed path. In QCD, the Wilson line is defined as [14]

$$U_P(x,y) = P[e^{ig\int_y^x A_\mu(z)dz^\mu}] \;, \qquad (4)$$

where symbol $P$ is responsible for the *path ordering* due to the noncommutativity of different $A_\mu$ in a non-Abelian field. Except for the noncommutativity of different parts of the path, Eq. (4) is usually *path-dependent*, just as in QED in the presence of a uniform magnetic field. Thus we use a superscript $s$ to denote a particular path. The differential equation of $U_P(x,y)$ that holds for QED also works for QCD along any path,[24]

$$D_\mu^s U_P(x,y) = 0 \;. \qquad (5)$$

The denotation $s$ here indicates that the differential is along an arbitrary path $s$. With Eq. (5), it is straightforward to verify that the Wilson line Eq. (4) acts as an evolution factor of the wave function for massless quarks

$$\psi_P^s(x,x_0) = U_P^s(x,x_0)\psi(x_0) \;, \qquad (6)$$

where $\psi(x,x_0)$ is the solution of the Dirac Equation

$$(i\partial\!\!\!/ - gA)\psi(x,x_0) = 0 \;, \qquad (7)$$

and $\psi(x_0)$ satisfies $(i\partial\!\!\!/ - m)\psi(x_0) = 0$ with $m = 0$. If $m \neq 0$ in Eq. (7), the mass term would contribute nothing to the phase factor when the path $x_0 \to x\,(x_0)$ forms a loop. In view of this fact, in what follows we take $U_P(x,y)$ as evolution factor of wave function also for massive quarks when we deal with closed paths.

A loop (closed path)

$$U_P(x,x) = P[e^{ig\oint A_\mu(z)dz^\mu}] \qquad (8)$$

defined on the basis of Eq.(4) in a non-Abelian field is not gauge invariant. So, in order to construct a gauge invariant Lagrangian of gauge field, the trace of Eq. (8) is performed to find a gauge invariant quantity. Such a quantity is defined as Wilson loop, from which the effective interacting potential between heavy quarks can be obtained[18, 19, 20]. Here we will not compute the trace but instead focus on Eq. (8) to investigate the topological property of wave function of Eq. (6). Applying Stokes's theorem to a loop, we can rewrite Eq. (8) as

$$U_P(x,x) = e^{i\frac{g}{2}\int_\Sigma F_{\mu\nu}(z)d\sigma^{\mu\nu}}, \qquad (9)$$

where $\Sigma$ is infinitesimal surface inside the closed loop $P$, $d\sigma^{\mu\nu}$ is an area element on this surface, and $F_{\mu\nu}$ is the field strength defined in Eq. (2). Note that here in our selected plane, $F_{\mu\nu}$ reduces to the form shown in Eq. (3). Thus the phase factor in Eq. (9) has a form of $\int_\Sigma \vec{B} \cdot d\vec{S}$, where $d\vec{S}$ now denotes the area element in place of $d\sigma^{\mu\nu}$. Similar to the Aharonov-Bohm phase in QED, now the Eq. (9) can be written as

$$U_P(x,x) = e^{i\frac{g}{2}\phi_0^a \lambda^a}, \qquad (10)$$

where $\phi_0^a = F_{xy}^a \varepsilon^2$ is the color flux, $\varepsilon^2$ denotes the infinitesimal area enclosed in the loop.

To quantitatively discuss the Aharonov-Bohm phase of a non-Abelian field, let's define a unit of $\phi_0^a$. Let $\Omega(x) = \exp[i\alpha^a(x)\frac{\lambda^a}{2}]$ denote an infinitesimal transformation near unit, the gauge transformation of vector field $A_\mu$ is $A_\mu \to \Omega(x)(A_\mu + \frac{1}{g}\partial_\mu)\Omega^\dagger(x)$, which can be derived from the corresponding transformation for Wilson line $U(x+\varepsilon, x) \to \Omega(x+\varepsilon)U(x+\varepsilon, x)\Omega^\dagger(x)$ by collecting coefficient proportional to $\varepsilon$. The transformation for Wilson line $U(x+\varepsilon, x)$ originates from the gauge transformation of Eq. (6). Now let's consider the integral in Eq. (8) after performing a gauge transformation to $A_\mu$. The original gauge field $A_\mu(x)$ is regular, so is $\Omega(x)A_\mu(x)\Omega^\dagger(x)$. Whereas the term $\Omega(x)\frac{1}{g}\partial_\mu \Omega^\dagger(x)$ becomes singular for it can be rewritten as $-\frac{1}{g}\partial_\mu \ln\Omega(x)$ by using $\partial_\mu(\Omega\Omega^\dagger) = 0$. The analysis suggests that in the integral of Eq. (8) only the singular term $\Omega(x)\frac{1}{g}\partial_\mu\Omega^\dagger(x)$ remains. Expanding the transformation of $A_\mu$ and collecting coefficient proportional to a generator $\lambda^a$, one has

$$A_\mu^a \to A_\mu^a + \frac{1}{g}\partial_\mu \alpha^a + f^{abc}A_\mu^b \alpha^c. \qquad (11)$$

Now it is natural for us to take $\frac{1}{g}$ as the unit of $\phi_0^a$, because the phase factor of Eq. (8) can be rewritten as an integral with integrand $\frac{1}{g}\frac{\partial_\mu \alpha^a \lambda^a}{2}$,

$$e^{ig\oint \frac{1}{g}\partial_\mu \alpha^a \frac{\lambda^a}{2} dx^\mu} = e^{i\oint \frac{\lambda^a}{2} d\alpha^a}, \qquad (12)$$

in which $\alpha^a$ gains formally an identity as winding angle.

In our case, the integral path in evolution factor of Eq. (6) should form closed path, i.e., $\psi_P^s(x,x_0) = U_P^s(x,x)\psi(x,x_0)$. To make definitions of evolution factor $U_P^s(x+\varepsilon, x)$ and gauge infinitesimal transformation $\Omega(x)$ for a wave function $\psi(x,x_0)$ consistent along a definite closed path, the $U_P^s(x,y)$ is required to be single-valued along the closed path. We will show that the equation (6) can be consistent if the topology of evolution factor mimics that of Aharonov-Bohm effect. Let's put a particular color to the *non-local quark* in the plane, then the symmetry of the vector field $A_\mu$ produced by the *di-quark* would be $SU(2)$, and the group generators in phase factor reduce from $SU(3)$ to Pauli matrices $\tau^i$. For a closed path, the evolution factor in Eq. (12) would be

$$e^{i\oint \frac{\lambda^a}{2} d\alpha^a} \to e^{i\oint \frac{\tau^k}{2} d\alpha^k} \to e^{i\pi \vec{n}\cdot \vec{\tau}}, \qquad (13)$$

where $\vec{n} = (n_1, n_2, n_3)$, $n_i$ is real for any given $i(=1,2,3)$. By applying formula $e^{i\vec{A}\cdot\vec{\tau}} = \cos A + i\frac{\vec{A}\cdot\vec{\tau}}{A}\sin A$ (in which $A = |\vec{A}|$), it is noticed that $e^{i\pi\vec{n}\cdot\vec{\tau}} = 1$ only when $|\vec{n}| = 2k$ ($k$-integer). An equivalent but easier way to understand the result is making only one of $n_i$ nontrivial, e.g. $n_1 = n_2 = 0$ and $n_3 = 2k$, while the physical meaning of $e^{i\pi \vec{n}\cdot\vec{\tau}}$ would not change with this particular choice. To put it another way, only when the *non-local quark* winds even loops does the nonintegrable phase Eq. (13) have the same meaning as Aharonov-Bohm effect and at the same time is the evolution factor of Eq. (6) for closed paths being able to be defined uniquely and consistently. In a geometrical picture, the *non-local quark* moves around a flux tube of two units, $\phi_{xy}^a = 2\phi_0^a$, but in fact it winds along the closed edge of a Möbius-strip[25] around a flux tube with only one unit. So in the language of fractional charge[26], the non-local quark is bound to a flux tube $2\phi_0^a$ to form a "composite particle". In the topological scenario [15] for FQHE, we find the filling factor $1/(2m+1/p)$ can fit our model with $m = 1$ and $p = 1$, hence a filling factor $1/3$ for the *non-local quark*.

In FQHE with filling factor $1/3$, an energy band is already full when the number of accommodated electrons reaches one third of the number of total states and any further filling to the same band is forbidden. If we view the three states in a baryon as a degenerate energy band, a quark with a filling factor $1/3$ can in fact fill all these three energy levels. This means that one quark has effectively occupied three states and any further filling by quarks with the same color is not allowed. Nevertheless, since all the three binding quarks share the *SU (3)* symmetry, the filling factor for one defined *non-local quark* is simultaneously applicable for the other two quarks. As a result, every single of the three quarks occupies three states (See Fig. 1). We cannot diagonalize the quarks' states by conventional transformation between representations. One can see quarks (states) as a whole with three colors, but cannot set them apart to get any one of them (See Fig. 2). Note that we initially assigned a particular color to the *non-local quark*, but finally the *non-local quark* loses its identity. The reason is that its color has been separated into three states, as illustrated in Fig. 2.

Similarly, the Hamiltonian method used in the study of FQHE can also be applied to quarks. If the color of the *non-local quark* has been specified, the Dirac Equation (7) will reduce to the same form as $(i\partial - gA - m)\psi(x, x_0) = 0$ with only two colors appearing in the wave function $\psi^{Transpose} = (\psi^a, \psi^b)$ ($a, b$ denote the un-specified colors). For heavy quarks, the approximation of small quantity $|\vec{p}|/m_{quark}$ is always available. Since in our FQHE gauge $A_\mu$ is projected onto a plane and its scalar part is vanishing, it can be shown following the standard approximation procedure in QED[27] that the large components of wave functions $\psi^a, \psi^b$ satisfy the following Schrödinger-Pauli equation,

$$i\frac{\partial}{\partial t}\psi = -\frac{1}{2m}[\vec{\sigma}\cdot(\vec{\nabla} - ig\vec{A})]^2\psi. \quad (14)$$

Here, the wave function $\psi$ has the same form as $\psi^{Transpose} = (\psi^a, \psi^b)$, but only the first two spinor components of each color remain; $\sigma^i$ are Pauli matrices for spins; and $A_{x,y} = A^a_{x,y}\tau^a$. Additionally, in FQHE the spin effect in Hamiltonian is suppressed,[28] so in heavy quark limit it is reasonable to assume that both spin components satisfy the same equation

$$i\frac{\partial}{\partial t}\psi^{1,2} = H\psi^{1,2}, \quad H = -\frac{1}{2m}(\vec{\nabla} - ig\vec{A})^2, \quad (15)$$

where $\psi^{a,b} = [(\psi^1, \psi^2)_{a,b}]^{transpose}$. This Hamiltonian is just the same as that used in FQHE, from which the energy gap for the ground state can be evaluated as $\Delta \sim \frac{e^2}{l_B}$, where $l_B$ is magnetic length defined by $l_B^2 = \left|\frac{1}{q_e B}\right|$ (Ref. [28]). In the case of quarks, the energy gap for the ground state is $\Delta \sim \frac{g^2}{l_B^a}$, with $(l_B^a)^2 = \left|\frac{1}{q_g F_{\mu\nu}^a}\right|$. This means that that the field induced by the *di-quark* determines the energy gap for the three-quark-system.

The analogy between three-quark baryon states and the FHQE studied in this work is a natural extension to previous assumptions and perceptions [29][30][31]. We expect this work to open new windows for further investigation on nonperturbative properties of quarks. Here, we have dealt with the non-local quark and its planar motion under the heavy-quark approximation $|\vec{p}|/m_{quark} \ll 1$. That ensures the credibility of the model and its relation to QCD is comparable with quark potential models.

Our model is valid for the non-Abelian field with symmetric group $SU(3)$, but not for any other $SU(N)$ fields of strong interactions. In this sense, the FQHE appeared in 2-dimension artificially designed condensed matter seems to be an intrinsic characteristic for quarks. The difference is that the FQHE in condensed matter is a cooperation involving a great number of particles whereas a hadron contains only a few constituent quarks. We stress that the correlation effect is a crucial point to understand the tri-quark confinement. Along this way, the implication of other fractional states of FQHE will be instructive in understanding quark states of other possible types. And applying the chiral-symmetry of 2-dimension Graphene[32] to light hadrons may shed light on the nonperturbative properties of light quarks, which are not involved in this work.

H.J.W. is grateful to Prof. S. S. Wu and Prof. Yu-Xin Liu for encouraging discussions.

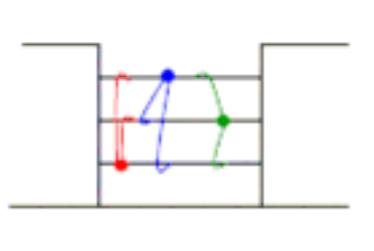

**FIG. 1. In a three-quark color-singlet state, every single of the three quarks occupies three states.**

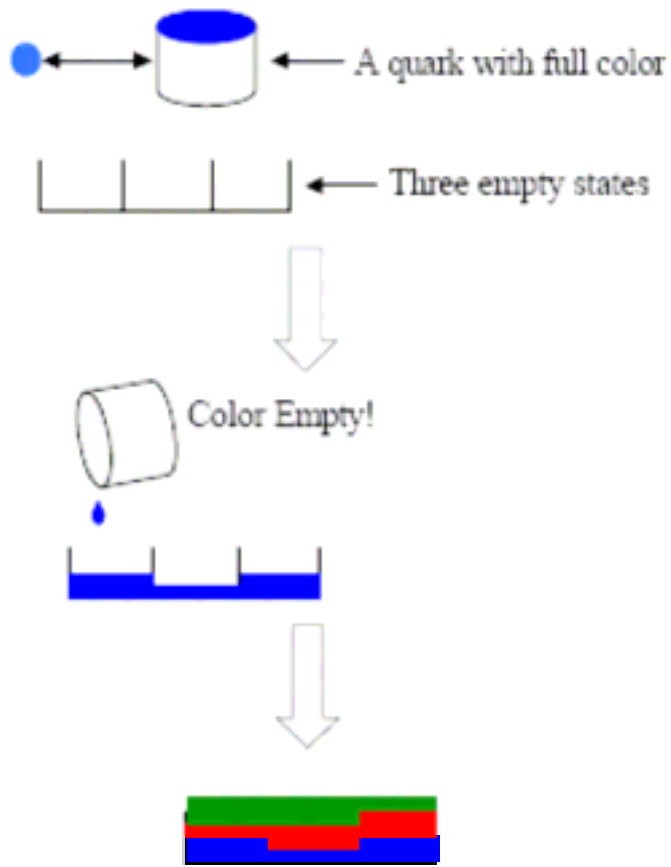

**FIG. 2. In a three-quark color-singlet state, the three quarks are viewed as a whole with three colors, but not able to be set apart.**